# Electromagnetic interactions derived from potentials: charge and magnetic dipole

Roberto Coïsson
Università di Parma, Dipartimento di Fisica e Scienze della terra
43100 Parma, Italy
e-mail roberto.coisson@fis.unipr.it

**Introduction**

In the teaching and applications of electromagnetism, the scalar potential φ is used in electrostatics as much and even more that the electric field **E**, as calculations are usually simpler with a scalar. For electromagnetism with slowly-moving charges and magnets, the vector potential **A** is used much less, in part probably because of an old prejudice, that it is only a mathemathical trick without "physical meaning", while, as it has been summarized in a recent review [1] (see also [2-5]), several papers along the history of electromagnetic studies have shown that, while φ is a potential energy (per unit charge), **A** can be interpreted as a potential momentum.

We want here to discuss a well-known [6] elementary case: the interaction (in vacuo) between a point charge and a magnetic dipole (MD), using potentials, along the lines indicated by Konopinsky [2], and to show that the "physical interpretation" might look quite different than using fields. In particular Lorentz's force does not appear explicitly, and has different interpretations in different cases. Also, the scalar potential from a moving MD appears as the potential from an equivalent electric dipole, and the electromagnetic momentum of the dipole in an electric field is a consequence of the mass-energy relationship.

**Basic formulae**

Let us recall some basic formulas for point particles, neglecting $v^2/c^2$:

Potentials of a point charge q at rest, at distance $\vec{r} = r\hat{r}$ :

$$\phi = \frac{1}{(4\pi\epsilon_0)}\frac{q}{r} \qquad \vec{A} = 0 \tag{1}$$

Potentials from a stationary magnetic dipole **m**:

$$\vec{A} = \frac{\mu_0}{4\pi}\frac{(\vec{m}\wedge\hat{r})}{r^2} \qquad \varphi = 0 \tag{2}$$

Potentials $\vec{A}'$, $\phi'$ as seen from an observer moving with velocity -**v** with repect to the particles, or an observer at rest while the particles are moving with velocity +**v** can be written, in the "galilean approximations" $v^2/c^2 \ll 1$ and negligible acceleration [7,8]:

moving charge: $\qquad \vec{A} = \frac{\vec{v}}{c^2}\varphi = \frac{\mu_0}{4\pi}\frac{q}{r}\vec{v} \qquad \varphi' = \varphi \tag{3}$

moving magn dipole $\qquad \varphi = \vec{v}\cdot\vec{A} = \frac{\mu_0}{4\pi}\vec{v}\cdot(\frac{\vec{m}\wedge\hat{r}}{r^2}) \qquad \vec{A}' = \vec{A} \tag{4}$

Eq. 4 can be interpreted as a scalar potential of an equivalent electric dipole $\vec{p} = \vec{v}\wedge\vec{m}/c^2$

Remark: when the particles move with constant velocity, the potentials and the fields appear to come from the present position of the particle, as if there was no retardation, as can be seen from the expression of the Liénard-Wiechert potentials (see for ex. [9]).

The electromagnetic interaction momentum $\vec{Q} = \int \vec{E}\wedge\vec{B}\, dV$ of the system point charge – magnetic dipole can be expressed, according to Furry [10] either as:

$$\vec{Q} = q\vec{A} \tag{5}$$

or, equivalently, from the MD at rest in the electrostatic field $\qquad \vec{Q}_{md} = \frac{1}{c^2}\vec{m}\wedge\vec{\nabla}\varphi \tag{7}$

If the charge q is moving with velocity v, $\vec{Q} = q\vec{A} + \frac{q\vec{v}}{c^2}\varphi$ (the last term can be interpreted as

mass-energy equivalence: energy/c² times velocity)

From eqs. (3,4) we can infer that the interaction energy $U=\int \vec{B}_1 \cdot \vec{B}_2 \, dV$ is $U=-q\vec{v}\cdot\vec{A}$ (v is the velocity of the charge, if the dipole is at rest) or $U=-\vec{m}\cdot(\vec{\nabla}\wedge\vec{A}\,')=\frac{-1}{c^2}\vec{m}\cdot(\vec{\nabla}\wedge\varphi\vec{v})$ .

From the formula of Konopinsky [2], the equation of motion of a point charge is:

$$\frac{d}{dt}[m\vec{v}+q\vec{A}]=-q\vec{\nabla}[\phi-\vec{v}\cdot\vec{A}] \tag{8}$$

writing the total derivative of **A** as sum of partial derivative and the convective term, if the acceleration is negligible (or the velocity is kept constant by an equal and opposite force) the electromagnetic force F can be written as a sum of an "inductive" term and an "electrostatic" term:

$$m\dot{\vec{v}}+q[\frac{\partial \vec{A}}{\partial t}+(\vec{v}\cdot\vec{\nabla})\vec{A}]=-q\vec{\nabla}[\varphi-\vec{v}\cdot\vec{A}] \tag{9}$$

or $\quad \vec{F}=-q[\frac{d\vec{A}\,'}{dt}+\vec{\nabla}\varphi\,'] \tag{10}$

where the prime indicates the values as seen from the moving particle.

**Examples**
The interaction of a magnetic dipole with a point charge is a simple example illustrating the use of potentials for calculating the force, momentum and angular momentum on an e.m. system.

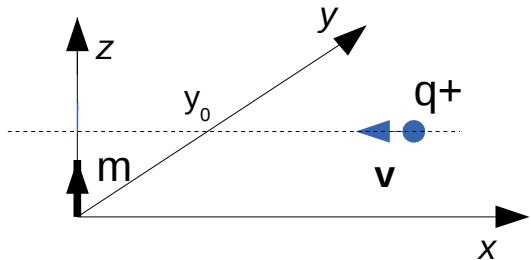

Figure 1: charge moving in the plane x-y along y= y₀

1) Force from a magn dip $\vec{m}$ at the origin and directed along +z on a charge q in the position (x,y₀) moving with velocity $-\hat{x}v$ (see fig.1): according to eq. 10 or 9, with φ=0 and $\partial A/\partial t=0$
(by $A_{i/j}$ we mean the derivative of i-th component of A with respect to the j-th direction), the force is

$$\vec{F}=q[(\vec{v}\cdot\vec{\nabla})\vec{A}-\vec{\nabla}(\vec{v}\cdot\vec{A})]=q[\hat{x}(A_{x/x}-A_{x/x})+\hat{y}(A_{x/y}-A_{y/x})]=\frac{-\mu_0}{4\pi}\frac{qmv\,\hat{y}}{(x^2+y_0^2)^{3/2}}(1+\frac{xy_0}{(x^2+y_0^2)}) \tag{12}$$

, when it arrives at x=0, the field momentum is q**A** , which is directed along -x. However it had been subjected to a force always perpendicular to x. In fact this force is the sum of two terms, dA'/dt and gradφ'. The first term, that we may call "induction" force, integrated over time, gives q**A** (and then contributes to the volume integral of the interaction momentum $\vec{E}_q\wedge\vec{B}_m$ ), while the other (that we may call "electrostatic" force), integrated over dx, gives $qAv/c^2$ which can be interpreted as mass (energy/c²) times velocity. This interaction mass/energy is the integral of $\vec{B}_q\cdot\vec{B}_m$ and is zero if the charge is at rest, in which case also the corresponding interaction energy is zero.
As we see, the contribution of the "inductive" and "electrostatic" terms vary with the position. Their resultant is the Lorentz force, perpendicular to the charge velocity (see fig. 2).
When the charge arrives at position (0,y) with velocity v=(-v,0) it has acquired momentum (in a direction perpendicular to the force that it had experienced!) and lost energy. This is true even if we have, instead of a dipole, an infinite solenoid and there is no magnetic field and no force.

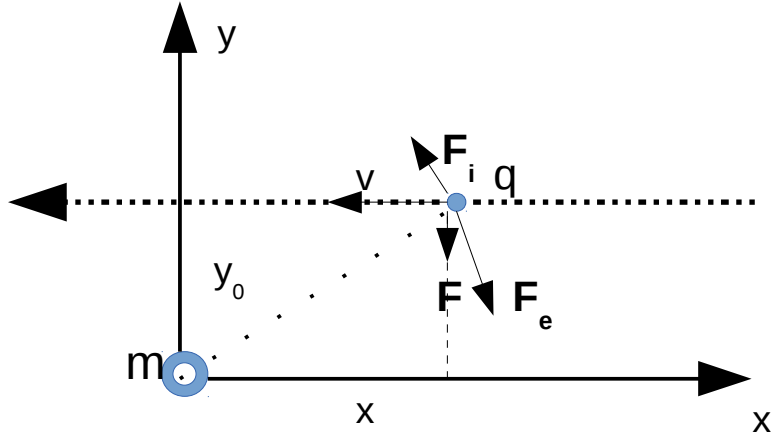

Figure 2: Forces acting on the charge q: (x,y$_0$) position of q, v its velocity, $F_i = q\, dA'/dt = q(\vec{v}\cdot\vec{\nabla})\vec{A}$ "induction" force, $F_e = q\nabla\varphi' = -q\nabla(\vec{v}\cdot\vec{A})$ "electrostatic" force, $F = F_i + F_e$ resultant Lorentz force, perpendicular to **v**.

We just mention that if y$_0$=0,
$$\vec{F} = -q[(\vec{v}\cdot\vec{\nabla})\vec{A} - \vec{\nabla}(\vec{v}\cdot\vec{A})] = \frac{-\mu_0}{2\pi}\frac{qm}{x^3}v\hat{y} + \frac{\mu_0}{4\pi}\frac{qm}{x^3}v\hat{y} = \frac{-\mu_0}{4\pi}\frac{qm}{x^3}v\hat{y} \qquad (11)$$
the two forces are in opposite direction and one is double than the other.
Of course, if A is irrotational, $(\vec{v}\cdot\vec{\nabla})\vec{A} - \vec{\nabla}(\vec{v}\cdot\vec{A}) = 0$ (no magnetic field).

2) Same example (with y$_0$=0), but both particles at rest. The momentum of the e.m. field of the system can be calculated as $\quad Q = qA = \frac{\mu_0}{4\pi}\frac{qm}{x^2}$ . $\qquad (13)$

The same result is obtained from the dipole (absolute value, taking into account that the magnetic moment is perpendicular to the potential gradient):
$$Q_{md} = \frac{1}{2c^2}m\nabla\phi = \frac{\mu_0}{4\pi}\frac{mq}{x^2} \qquad (14)$$
(this can be interpreted as a momentum due to mass-energy equivalence as component charges move in different values of φ: the momentum has the direction of the velocity of the charges moving in a higher potential, as can be shown by imagining the dipole as a square loop of current)

But if we imagine to move the charge from infinity to position x, and we integrate the force necessary to keep it "on rail" in order to obtain the impulse given by it to the charge, we obtain half that value:
$$\int_{-\infty}^{x} F\, dt = \frac{-\mu_0 mq}{4\pi}\int_{\infty}^{x} x^{-3}\, dx = \frac{\mu_0 qm}{8\pi x^2} \qquad (15)$$

The situation is similar to the case of energy of the e.m. field of two point charges q$_1$ and q$_2$, However in that case the work done on one charge in moving it from infinity to the position x is equal to the whole energy of the fields, as the force necessary to keep the other charge at rest does not produce work.
But there is a problem: in this case the total electromagnetic interaction momentum qA is the double of what we have just calculated; that means that the time integral of the force keeping the dipole at rest must also be equal to qA/2, and must have the same sign. This seems in contradiction with the fact that actually the force on the dipole has a direction opposite to the force on the charge (for momentum conservation).
We have to consider that the whole system charge + magnetic dipole both at rest, has an electromagnetic interaction momentum in the direction +y, which is compensated by a non-e.m. contribution called
"hidden momentum" [11,1]: $\quad Q_H = \frac{-1}{c^2}m\nabla\phi = \frac{-\mu_0}{4\pi}\frac{mq}{x^2} \qquad (16)$

This term is independent of the model we make for the MD, but its "physical interpretation" depends on the model: in the free charge case, it is due to the relativistic mass increase due to velocity, in the case of frozen charges on a dielectric it corresponds to an energy flow due to stresses in the supporting disk: in any case it is due to the trasformation of the momentum-energy-stress-kinetic tensor [12-14].

In conclusion, when the system is put together by bringing the charge from infinity, on the dipole there are two forces: one qdA/2dt (electromagnetic) in the same direction as the force on the charge, and one non-electromagnetic (could be called mechanical, relativistic), the time derivative of the hidden momentum, double of the previous value and with opposite direction, so the resultant is equal and opposite to the force on the charge.

If we make the same integral as above but for the angular momentum with respect to the origin (where the magnetic dipole is located), we get, as can be expected,

$$\int_{-\infty}^{x} F\, x\, dt = \frac{-\mu_0 m q}{4\pi} \int_{\infty}^{x} x^{-2} dx = \frac{\mu_0 q m}{4\pi x} = q A x \qquad (17)$$

3) If both particles are moving together, the two transformations of the potentials (motion of the source and of the other particle) compensate each other.
We can remark here that it comes out quite naturally that if the force is zero in the rest system it is zero also in moving systems, while this is not so evident when we calculate forces using the fields [15].

**Acknowledgement**: this note has profited from discussions with, and revision by, Giovanni Asti and from comments by Germain Rousseaux.